 \documentclass[11pt,a4paper]{article}
 \usepackage{amsmath}
\usepackage{cite}
\usepackage{graphicx}
\usepackage{amsfonts}
\usepackage{amssymb}
\input epsf

\def\m{\mu}
\def\n{\nu}
\def\a{\alpha}
\def\b{\beta}

\def\de{\delta}

\def\la{\lambda}
\def\La{\Lambda}

\def\ka{\varkappa}

\def\vro{\varrho}
\def\ga{\gamma}

\def\sh{{\mathrm{sh}}}
\def\ch{{\mathrm{ch}}}

\def\arctg{{\mathrm{arc \, tg}}}

\def\arccoth{{\mathrm{arc \, cth}}}

\def\coth{{\mathrm{cth}}}

\def\Hyp{{}_2{\mathrm{F}}_1}
\def\Bet{{\mathrm{B}}}

\def\dd{{\mathrm{d}}}

\def\acc{{\mathrm{acc}}}

\def\eff{{\mathrm{eff}}}
\def\GBL{{\mathrm{GBL}}}

\def\GB{{\mathrm{GB}}}
\def\QTG{{\mathrm{QTG}}}
\def\MS{{\mathrm{MS}}}
\def\KH{{\mathrm{KH}}}
\def\KW{{\mathrm{KW}}}
\def\cd{{\mathrm{c.d.}}}
\def\edge{{\mathrm{edge}}}

\def\Pl{{\mathrm{Pl}}}

\def\dS{{\mathrm{dS}}}

\def\eff{{\mathrm{eff}}}

\def\ini{{\mathrm{initial}}}

\def\rm{\mathrm}
\def\cal{\mathcal}

\def\pa{\partial}

\def\be{\begin{equation}}
\def\ee{\end{equation}}
\def\br{\begin{eqnarray}}
\def\er{\end{eqnarray}}
\def\bsub{\begin{subequations}}
\def\esub{\end{subequations}}

\begin{document}




\begin{titlepage}
\vspace*{-2 cm}
\noindent
\begin{flushright}
\end{flushright}

\vskip 1 cm
\begin{center}
{\Large\bf Energy Density Bounds in Cubic Quasi-Topological Cosmology} \vglue 1  true cm

  {U. Camara dS}$^{*}$\footnote {e-mail: ulyssescamara@gmail.com}, { A.A. Lima}$^{*}$\footnote {e-mail: andrealves.fis@gmail.com} and { G.M.Sotkov}$^{*}$\footnote {e-mail: sotkov@cce.ufes.br, gsotkov@yahoo.com.br}\\

\vspace{1 cm}

${}^*\;${\footnotesize Departamento de F\'isica - CCE\\
Universidade Federal do Esp\'irito Santo\\
29075-900, Vit\'oria - ES, Brazil}\\

\vspace{5 cm}

\end{center}

\normalsize
\vskip 0.5cm

\begin{center}
{ {\bf ABSTRACT}}\\
\end{center}

\vspace{0.5cm}

We investigate the thermodynamical and causal consistency of cosmological
models of the cubic Quasi-Topological Gravity (QTG) in four
dimensions, as well as their phenomenological consequences. 
Specific restrictions on the maximal values of the matter densities are derived  by  requiring  the apparent horizon's entropy  to be  a non-negative, non-decreasing function of time.  
The QTG  counterpart  of the Einstein-Hilbert (EH) gravity model of linear  equation of state is
studied in detail. An important feature of this particular QTG cosmological model is the new early-time acceleration period  of the
evolution of the Universe, together with the standard late-time acceleration present in the original EH model. The QTG correction to the causal diamond's volume is also calculated.

\vspace{0.5 cm} 

\end{titlepage}

\tableofcontents 


\section{Introduction}
\label{Introduction}

Regions containing extremely dense matter are a common feature of all the big-bang inflationary cosmological space-times. The consistent description of such high energy states in the evolution of the Universe  requires certain ``higher curvature'' extensions of Einstein-Hilbert (EH) gravity, involving powers (of the traces) of the Riemann tensor, believed to take into account the short distance quantum effects. Such corrections to the EH action are known to arise as counter-terms in the  \emph{perturbative}  quantization  both of matter in curved spaces \cite{BirrelDavies} and of pure EH gravity  as well \cite{'tHooft:1974bx, Stelle:1976gc}. The main problem with ``higher curvature'' gravity theories concerns the presence of higher derivatives of the metric in the equations of motion, which in general lead to causal and unitarity inconsistencies \cite{'tHooft:1974bx, Stelle:1976gc} when considered out of the framework of  superstring theory. Nevertheless, the extensive studies of such models, in particular the so called modified $f(R)$-gravities, have found  interesting applications in different areas of  modern cosmology (see, e.g., \cite{Nojiri:2010wj, Capozziello:2011et, Clifton:2011jh} and references therein).

The present paper is devoted to the investigation of the effects caused by the higher curvature terms in a particular cosmological model in  four dimensions, based on the  simplest ``most physical'' extended gravity, given by the following  cubic action for Quasi-Topological Gravity  \cite{Oliva-Ray1}:
\br
&\hspace{-1cm} S_\GBL = \int\! \frac{\sqrt{-g} \, \dd^4x}{\ka^2} \Big\{ R - \la L^2 \left[ 2 R^2 - 8 R_{\a\b} R^{\a\b} + 2 R_{\a\b\m\n}R^{\a\b\m\n} \right]\nonumber \\
&\hspace{-1cm} + \frac{\m L^4}{4} \Big[R^3 +  18 R^{\a\b\ga\de}R_{\ga\de\m\n}R^{\m\n}{}{}_{\a\b} - 40 R^{\a\b}_{}{}_{\ga\de}R^{\ga\m}{}{}_{\b\n}R^{\de\n}_{}{}_{\a\m} \nonumber \\
&\hspace{-1cm}-36 R^{\a\b\ga\de}R_{\a\ga}R_{\b\de} + 8 R^{\a\b}R_{\b\ga}R^{\ga}{}_{\a}  \big] + {\cal L}_{\rm{matter}}  \big\},\label{action non cf}
\er
where $\ka^2 \equiv 16 \pi G = 4 l_{\Pl}^2$; $l_\Pl$ is the Planck length; $\la$, $\m$ are dimensionless ``gravitational coupling'' constants, while $L$ is a new length scale, which can be chosen as  $L = l_\Pl$ by an appropriate redefinition of $\la$ and $\mu$. 
We include the cosmological constant $\Lambda_0>0$ implicitly in the matter Lagrangian ${\cal L}_{\rm{matter}}$. 

The remarkable feature of this cubic extension of the EH  action, is that the corresponding equations of motion for all conformally flat metrics, as for example those of domain walls  and of flat Friedmann-Robertson-Walker (FRW) space-times:
\br
\dd s^2 = - \dd t^2 + a^2(t) \, \dd x^i \dd x_i \; ,  \label{FRW metric}
\er
 are of \emph{second order} \cite{Nosso_Paper}, while arbitrary metrics yield, in general, equations of motion of fourth order\footnote{ 
The particular combination of quadratic terms represents the $d=4$ Gauss-Bonnet topological invariant, and does not contribute to the dynamics.}. This fact, together with the introduction of an appropriate superpotential  and the related BPS-like first order system  of equations for the cubic QTG-matter model (\ref{action non cf}), derived in Ref.\cite{Nosso_Paper}, provide an efficient method for the analytic construction of a large family of exact flat FRW's solutions, representing asymptotically $\dS_4$ space-times.

The most important and universal new property of  these higher curvature cosmological models  is that, differently from the EH case, the entropy  prescribed to  the apparent horizons \cite{Sinha_cosmo}
\begin{eqnarray}
&  s(t) = \frac{16 \pi^2}{\ka^2} \, \frac{1}{ H^{2}} \left( 1 - 2 \la L^2 \,  H^2 + 3 \m L^4 \, H^4  \right)  \,  \label{s(H)} 
\end{eqnarray}  
with $H \equiv \dot a / a$ denoting the Hubble factor, \emph{is not automatically positive definite and increasing}. As a consequence, the requirement of the  thermodynamical  consistency of these models: $s(t)\geq 0$ and $\dd s/ \dd t > 0$, introduces certain restrictions on the available maximal values $\vro_{\max}$ of the matter densities $\vro = \frac{6}{\ka^2} H^2 \leq \vro_{\max}$ and certain minimal scales $L_{\min}$ (related to $\vro_{\max}$) up to which we can have a physically meaningful description of the Universe evolution within the framework of the cubic QTG cosmologies.

In order to exemplify the effects caused by the higher curvature terms, we choose a particularly  simple and yet quite rich cosmological model, representing a QTG extension of the following EH model of a \emph{linear equation of state}:
\br
p_0 / \vro_0  = w_1 - w_2 / \vro_0 \; . \label{EoS exemplo}
\er
The matter content is that of a barotropic fluid with constant equation of state parameter $w_1$, together with a dark energy ``fluid'' representing the cosmological constant $\La_0 = \ka^2 w_2 / 2 (1 + w_1)$.
This EH cosmological model has been widely studied \cite{Bousso_hepth_0702115, Chavanis_astroph_1208.0801v1, KaloperLinde, Babichev_etal}, with a particularly remarkable result by Bousso et al. \cite{Bousso_hepth_0702115}, who deduce the value of the cosmological constant for a universe dominated by dust through most of its history ($w_1 = 0$). The new QTG  features established in the present paper are: The presence of a new (early-time) acceleration period; changes in the duration of the acceleration and deceleration periods, as well as of the  future and past event horizon radii; and finally certain very small corrections to the volume of the causal diamond (to be compared with the EH one \cite {Bousso_hepth_0702115}).


\section{Modified FRW Cosmology} \label{Modified FRW Cosmology}

Consider an universe filled with a barotropic fluid with energy density $\vro_0$ and pressure $p_0$,  components of a `bare' energy-momentum matter-tensor $T^{(0)}_{\m\n}$.
 For the ansatz (\ref{FRW metric}), the QTG equations of motion derived from (\ref{action non cf}) are the modified Friedmann equations:
\bsub\label{fluid eq QTG all}\br
&& \ka^2 \vro_0 = 6 H^2 \left( 1- \m L^4 \, H^4 \right) \; , \label{fluid eq QTG i}\\
&& \ka^2 (\vro_0 + p_0) = - 4 \dot H \left(1 - 3 \m L^4 \, H^4 \right) \; ; \label{fluid eq QTG} \\
&&\dot \vro_0 + 3 H (p_0 + \vro_0) = 0 \; . \label{cont 0}
\er\esub
They reduce to the usual EH-Friedmann equations when $\m = 0$; otherwise the contributions from the QTG terms may be regarded as separating a ``gravitational energy momentum tensor''
$T^{\QTG}_{\m\n}$ at the right-hand side of the Einstein equations, thus composing an \emph{effective energy momentum tensor} $T^\eff_{\m\n}$, viz.
\br
G_{\m\n} = T^\eff_{\m\n} \; ; \quad T^\eff_{\m\n} = T^{(0)}_{\m\n} + T^{\QTG}_{\m\n} ,	\nonumber
\er
where $G_{\m\n}$ is the Einstein tensor. 
Notice that the effective energy-momentum tensor has all the properties of a perfect fluid tensor, i.e. its components, $T^\eff_{\m\n} = {\rm{Diag}} \; (-\vro , p, p, p)$ satisfy the usual, EH Friedmann equations:
\br
& \vro = \frac{6}{\ka^2} H^2 , \quad  \vro + p = - \frac{4}{\ka^2} \dot H \; , \label{fluid eq eff}
\er 
as well as the continuity equation, 
\br
\dot \vro + 3 H (p + \vro) = 0 \; , \label{cont eff}
\er
which is a simple consequence of the Bianchi identities. 
Because of its direct connection to the Hubble function, the  cosmological observations (of distances and red shifts) should perceive not the bare energy density $\vro_0$, but rather the effective one, $\vro$, related to the former via Eq.(\ref{fluid eq QTG i}): 
\br
& \vro_0 = \vro \left( 1- \m L^4 \frac{\ka^4}{36} \vro^2 \right) \; . \label{vro 0 eff}
\er
Although such a hydrodynamical interpretation of $T^\eff_{\m\n}$  is rather formal, we further assume that this ``effective fluid'' obeys the weak energy condition,  i.e. $p + \vro \geq 0$, what assures that $\dot H \leq 0$. Then if the bare fluid also satisfies the weak energy condition, the function
\br
& C_0 \equiv 1 - 3 \m L^4 \, H^4 = 1 - \frac{1}{12}  \m L^4 \ka^4 \vro^2  \label{C0}
\er
must be positive (or vanishing), as can be seen from Eq.(\ref{fluid eq QTG}). 
While this condition holds  automatically for $\m \leq 0$, for positive values of $\m$ the sign of $C_0$ will depend on the value of the energy density; it vanishes for $\vro = \frac{2}{L^2 \ka^2} \sqrt{3 / \m}$ and it is indeed \emph{negative} for greater values of $\vro$. Since near a singularity the value of $\vro$ grows without bounds, for $\m > 0$ it is only possible that both the bare and the effective fluids satisfy the weak energy condition in a \emph{nonsingular} universe.  This would be the case, for example, in bounce-like models beginning and ending at de Sitter spaces with (asymptotic) densities $\vro_{{\rm{dS}}} \leq \frac{2}{L^2 \ka^2} \sqrt{3 / \m}$. We leave the study of such spaces for a more thorough discussion in \cite{LoveCosmo_tobepublished}, and focus for the remainder of this letter in the singular cases, thus considering only $\m \leq 0$.


\section{Horizon entropy} \label{sect. horizon entropy}

 Killing event horizons are known to posses thermodynamical  properties: a temperature $T$ related to the surface gravity, and an entropy $s$ which in EH gravity is given by the Bekenstein-Hawking formula $s = A / 4$ (in geometrized units), $A$ being the area of the horizon \cite{Bardeen:1973gs, Hawking:1974sw, Gibbons:1977mu}. Then the  Einstein equations can be rewritten as a Clausius relation \cite{Jacobson1}, 
$\dd E = T \, \dd s$, for the flux of energy $\dd E$ across the horizon%
\footnote{ Remarkably, this equivalence remains  valid  in a variety of ``higher curvature'' gravitational theories \cite{Jacobson2, Padma, Cai:2005ra, Akbar:2006kj} with the horizon's entropy given then by the Wald formula \cite{Wald:1993nt, Iyer:1994ys}.}.

The lack of time-like Killing vectors in non-stationary space-times -- as for example  typical FRW spaces  -- represents an obstacle in the definition of a surface gravity for the corresponding \emph{dynamical} apparent horizons. A possible consistent  generalization of the ``Thermodynamics/Gravity'' correspondence  can nevertheless be achieved by using the Kodama vector to define the horizon's Kodama-Hayward temperature $T_{\KH}$ \cite{Hayward}.  As it was recently shown by Cai et al. \cite{Cai:2005ra, Akbar:2006kj}, the corresponding Friedmann
equations, for EH gravity and for certain modified theories as well, turn out to be again equivalent to  the Clausius relation $\dd E_{\MS} = T_{\KH} \, \dd s_{\KW}$, with $E_{\MS}$ being the Misner-Sharp energy and $s_{\KW}$ an appropriately defined Kodama-Wald entropy, which for the cubic QTG cosmologies  is given by Eq.(\ref{s(H)}).%
\footnote{The proof of the equivalence between the QTG  modified Friedmann equations (\ref{fluid eq QTG all}) and the above generalization of the Clausius relation is given in our forthcoming paper \cite{alf}.}
 Notice that for de Sitter space-times, when $H$ is constant and the apparent horizon coincides with the Killing event horizon, our formula (\ref{s(H)}) reduces to the known static QTG  Wald entropy \cite{Sinha_cosmo}.
 
The consistent interpretation of $s_{\QTG}(t)$, given by Eq.(\ref{s(H)}), as an entropy  for the apparent horizon in the considered QTG cosmological models requires that it must be a non-negative and non-decreasing function of time. The later is always  true if both effective and bare fluids do satisfy the weak energy condition. Then as a consequence we  have that  $\dot s = - \frac{3}{2 G \ka^2} \, \frac{C_0(\vro)}{\vro^2} \, \dot \vro  \geq 0$. The restrictions imposed by the positivity condition $s_{\QTG}(t)\geq 0$ are slightly  more involved: depending on the signs and values of $\la$ and $\mu$, they  turn out to introduce certain upper bounds on the energy density $\vro$.


\noindent
\textbf{\textit{ Gauss-Bonnet Gravity:}} The topological nature of the GB term in the action renders the dynamics (i.e. the equations of motion) of GB gravity, for which $\m = 0$ and $\la \neq 0$, identical to that of EH gravity.  But the horizon entropy is \emph{not}  simply proportional to $H^{-2}$, in fact
\br
& s(t) = \frac{1}{4 G} \left(  \frac{1}{ H^{2}} - 2 \la L^2  \right) \; .  \nonumber
\er
Thus if $\la < 0$ the entropy density is always positive, but for $\la > 0$ there is a value of $H(t)$ -- or, equivalently, of $\vro$ -- past which the entropy becomes \emph{negative}. Therefore the assumption of positivity of entropy places  as an upper boundary, $\vro_\GB$,  on the possible values of the energy density:
\br
 0 \leq \vro \leq \vro_\GB \; ; \quad \vro_\GB \equiv \frac{3}{\ka^2 L^2 \la} \;  \label{vro GB}
\er
 \noindent
\textbf{\textit{Quasi-Topological Gravity:}}
Similar considerations applied to the QTG model, for  the negative values of the coupling $\m < 0$ we are interested in, lead us to the conclusion that the apparent horizon entropy is only positive for densities within the \emph{finite} interval: 
\br
0 \leq \vro \leq \vro_\QTG \equiv  \frac{2}{ \ka^2 L^2} \, \frac{1}{ \m} \left( \la  - \sqrt{\la^2 - 3 \m} \right) .\label{vro QTG}
\er
In a singular universe, it is then inevitable that at some instant the (divergent) effective energy density  violates the entropic threshold of Eqs.(\ref{vro GB}) or (\ref{vro QTG}) for a finite value of $\vro
$. Hence in the considered GB gravity (with $\la>0$) and QTG models of $\mu<0$, the cosmological singularity lies in a region of space-time which is already \emph{unphysical, for the apparent horizon entropy is negative.}


\section{An example of QTG cosmology}

In order to describe the changes in  the evolution of the homogeneous and isotropic Universe caused by the cubic QTG terms, we next address the problem concerning the construction of  analytic solutions of the modified  Friedmann  equations (\ref{fluid eq QTG all}) in the  particular example of the matter stress-energy tensor $T^{(0)}_{\m\n}$, whose components are related by the following \emph{linear} barotropic equation of state \cite{Bousso_hepth_0702115, Chavanis_astroph_1208.0801v1, KaloperLinde, Babichev_etal}: 
\br
p_0 / \vro_0 = \omega_0(\vro_0) \; ; \quad \omega_0(\vro_0) = w_1 - w_2 / \vro_0 \;. \label{EoS exemplo 2}
\er
 The constant $w_2$, if positive, represents an energy density which contributes to the pressure $p_0$ independently of the variable energy density $\vro_0$ -- thus when the former dominates (i.e. $w_2 / \vro_0 \gg 1$),  the dynamics is driven by a constant energy density with EoS $p_0 = - w_2$, resulting in a de Sitter geometry.  In this sense, $w_2$ may be identified with the cosmological constant and (\ref{EoS exemplo 2}) is an example of a simple  quintessence model.  The constant $w_1$, which may be seen as the ``matter EoS parameter'' (as opposed to the ``cosmological constant parameter'' $w_2$) is equal to the velocity of sound in the fluid: $v_0^2 = \pa p_0 / \pa \vro_0 = w_1$ (in Plank units, $c=1=\hbar$). The  causality condition $w_1 < 1$ excludes superluminal velocities, and if we also assume that the universe \emph{does not have a ``phantom'' phase}, then we have to impose $|w_1| < 1$.

The dynamics of the universe in such QTG cosmology may be found by solving the
 differential equation for $H(t)$  obtained by combining Eqs.(\ref{fluid eq QTG all}) with the EoS (\ref{EoS exemplo 2}):
\br
 \hspace{-0.7 cm} ( 1 - 3 \m L^4 H^4 ) \dot H = - \frac{3}{2} H^2 ( 1 - \m L^4 H^4 ) (1 + w_1) + \frac{\ka^2}{4} w_2 , \label{before expansion}
\er
the integration of which yields
\br
\hspace{-0.8 cm} &t(H) = - \frac{2}{3 (1+ w_1)} \frac{1}{L^4 \m}   \Bigg\{ 
2 \upsilon\left[  \arctg \left( \frac{H - \xi}{\zeta} \right) +\arctg \left( \frac{H + \xi}{\zeta} \right) \right] \nonumber\\
\hspace{-1.2 cm}&-\chi  \log \left( \frac{(H - \xi)^2  + \zeta^2}{(H + \xi)^2 + \zeta^2} \right) + 
\frac{\a}{H_\La} \, \arccoth \left( H / H_\La \right) 
- 2 \pi \upsilon \Bigg\} . \label{t(H)}
\er
Here $H_\La^2$ is the only real root of the cubic equation (\ref{vro 0 eff}) when written in terms of $H^2 = \ka^2 \vro / 6$ and solved for $H^2$, with $H_0^2 \equiv \ka^2 \vro_0 / 6$ considered known. The real and imaginary parts of the remaining roots, $h^2$ and its complex conjugate $\bar h^2$, are denoted by  $\xi$ and $\zeta$, while $\chi$ and $\upsilon$ are the real and imaginary parts of $\a /2 \equiv \frac{1 - 3 \m L^4 h^4}{2 h (h^2 - H_\La^2)(h^2 - \bar{h}^2)}$. Notice that we have chosen one specific singular solution $H (t \to 0) \to \infty$ of  Eq.(\ref{before expansion}), representing  big-bang space-times with singularity at $t=0$. Depending on the initial conditions imposed on $H(t)$ (or equivalently on $\vro(t)$) one can construct other non-singular bounce-like solutions (both for QTG or EH models), which are however out of the scope of the problems discussed in the present paper.

 The function $t(H)$ (\ref{t(H)}) is not trivially invertible; nevertheless it allows to describe the properties of the different periods of acceleration and deceleration of the QTG corrected Universe evolution, that can be obtained directly  from Eqs.(\ref{before expansion}), (\ref{fluid eq QTG}) and (\ref{fluid eq eff}). We next recall that one can also use the \emph{deceleration parameter} $q \equiv - \ddot a  \, a / \dot a^2$ (for $\dot a \neq 0$),  written in a suggestive form:
\br
& q = \frac{1}{2} (1 + 3 \, p / \vro ) \; ,
\er
in order to determine whether the universe undergoes  accelerated ($q < 0$) or  decelerated ($q > 0$) expansion. Thus, the ratio $p / \vro \equiv \omega_\eff$ between the components of the effective energy-momentum tensor plays the role of an `\emph{effective equation of state}'.%
\footnote{Changes in the EoS due to the higher curvature terms arising within the context of $f(R)$-modified gravity have been studied in Refs.\cite{Capozziello:2005mj, Capozziello:2005pa}.}
 Its explicit form :
\br
\omega_\eff (\vro) = - 1 + \frac{(1 + w_1) \left( 1 - \m L^4 \ka^4 \vro^2 / 36 \right) \vro -w_2}{ \vro \left( 1 - \m L^4 \ka^4 \vro^2/12 \right) } \;  \label{EoS eff}
\er
is derived by substituting Eq.(\ref{EoS exemplo 2}) into  Eqs.(\ref{fluid eq QTG}) and (\ref{fluid eq eff}). As expected, for $\m = 0$  we  get  $\omega_\eff = \omega_0$. Since $q \gtrless 0$ iff $\omega_\eff \gtrless -1/3$, it is convenient to consider only Eq.(\ref{EoS eff}). Observe that as one approaches the initial singularity and $\vro$ diverges, we can take the zeroth order limit of $1 / (\m L^4 \ka^4 \vro^2) \ll 1$ in order to show that $1 + \omega_\eff \approx \frac{1}{3} (1 + \omega_0)$. Therefore, even if $\omega_0$ is in the ``most decelerated range'' possible, viz. $\omega_0 \lesssim 1$, in the considered QTG cosmology we have an accelerated phase: $\omega_\eff \lesssim - 1/3$. Thus, the addition of the cubic QTG terms (\ref{action non cf}) to the EH action results in a \emph{new acceleration period at the beginning of the universe.}

\begin{figure}
\includegraphics{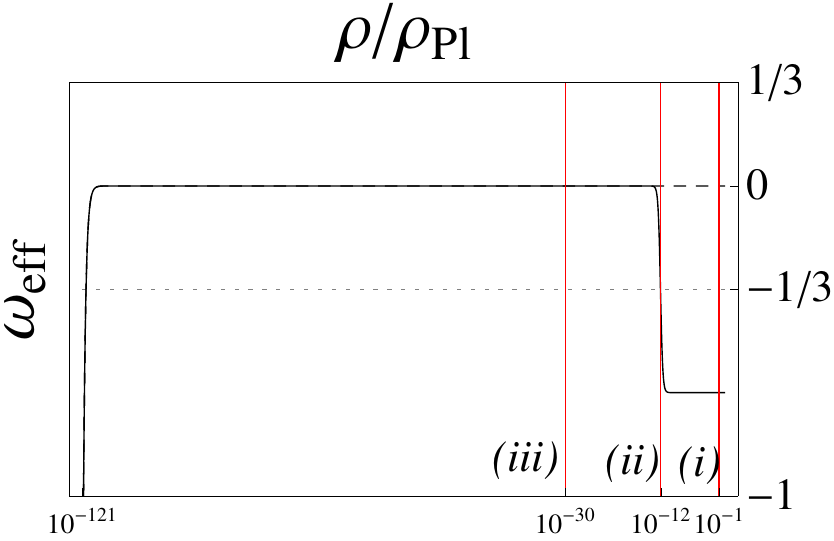}
\caption{ Evolution of the equation of state parameter for $\m = - 10^{24}$: 
the solid line corresponds to the effective EoS of QTG as a function of the effective energy density, Eq.(\ref{EoS eff}); the dashed black line corresponds to the EH case, $\omega_0(\vro_0)$; the red lines are the entropy densities of the apparent cosmological horizon, Eq.(\ref{s(H)}), for \textit{(i)} $\la = -9.00 \times 10^{22}$ ; \textit{(ii)} $\la = -2.50 \times 10^{51}$ ; \textit{(iii)} $\la = 7.53 \times 10^{69}$.}
\label{fig. EoS}
\end{figure}

This effect may be seen in Fig.\ref{fig. EoS}, where $\omega_0(\vro_0)$, given by Eq.(\ref{EoS exemplo 2}), is shown as the dashed line, while the continuous black line depicts the effective equation of state (\ref{EoS eff}). Here $w_1 = 0$, and $w_2$ is fixed by the observed value of the cosmological constant (cf. Eq.(\ref{La0 e w2})). One can clearly see that as the energy density increases the plot of $\omega_\eff$ sinks beneath the line of $-1/3$, indicating  the   new period of accelerated expansion. For smaller densities, however, there is very little difference between the EH plot and the QTG one.  Although some of the indispensable features of inflation -- slow-roll, for example -- \emph{are not present}, we shall  freely nominate this early acceleration period as an ``inflationary''.  The duration of  such a ``rustic inflation'' evidently depends on the value of $\m$. By imposing $\omega_\eff (\vro_\acc) = -1/3$, we get a cubic equation:
\br
& \frac{(w_1 - 1)}{36} \ka^4 \, \m L^4 \; \vro_\acc^3 - \frac{3 w_1 + 1}{3} \vro_\acc + w_2 = 0 \; , \label{cubic acc}
\er 
whose positive real roots give the threshold of the two acceleration periods now present in the dynamics -- the initial one  due to QTG and the final due to the cosmological constant.
The number of distinct real solutions depends on the sign of the discriminant of (\ref{cubic acc}) and, for $\m < 0$ and $w_1 > -1/3$, it determines a critical value
\br
 |\m_{{\rm{acc}}}| = \frac{16 (3 w_1 + 1)^3}{81 (1 - w_1) \ka^4 w_2^2 L^4} \nonumber
\er
for which the discriminant vanishes. Thus if $|\m| < |\m_{{\rm{acc}}}|$ there are two positive real roots for (\ref{cubic acc}), corresponding to two distinct periods of acceleration.
But for large enough values of the gravitational coupling, namely $|\m| \geq |\m_{{\rm{acc}}}|$, there is no positive real root to (\ref{cubic acc}) and as a consequence  the initial QTG acceleration period lasts for such a long time that it merges with the final one -- the universe is then never decelerated.

 It is worthwhile to remark here that in the QTG counterpart (\ref{EoS eff}) of the EH linear EoS, for $\m < 0$ the effective speed of sound $v_\eff^2 = \partial p / \partial \vro$   also satisfies  the  causality conditions $-1 < v_\eff^2 < 1$, under the same restrictions on the parameters $w_1$ and $w_2$.  On the other hand, for $\m > 0$, we see that $v_\eff^2 \to \infty$ as $\vro \to \frac{2}{L^2 \ka^2} \sqrt{3 / \m}$. This \emph{non-causal behaviour} of the effective fluid near the points where $C_0(\vro) = 0$ is yet another reason to consider here only the case of negative values of $\m$.

 Although the form of  $t(H)$ given by Eq.(\ref{t(H)}) does not allow to analytically determine the exact form of the Hubble function $H(t)$,
we may invert it in a first order approximation.  This is possible when the dimensionless quantity $|\m| L^4 H^4$ is much smaller than unity. For $L = l_\Pl$, such an approximation is valid during most of the universe history, since $H(t)$ is typically of a cosmological order (greater than 1 Mpc $\sim 10^{57} \times l_\Pl$). In what follows, we shall refer to this approximation as ``first order in $|\m|$'':
\br
H(t;\m) = H_0(t) + \m \, H_1(t) + \cdots ,\nonumber\\
 a(t,\m) = a_0(t) \left[ 1 + \m \, A_1(t) + \cdots \right],\nonumber
\er
 with the scale factor parametrized as  $a(t) = e^{A(t)}$. In zeroth order we have, naturally, the EH solution:
\br
& H_0(t) = \ka \sqrt{\frac{w_2}{6(1 + w_1)}} \, \coth \left[ (t - t_0) /  \tau  \right] ,\label{H0 t} \\
& a_0(t) =     \tilde a_0 \, \sh^{\de} \left[ (t - t_0) / \tau \right] , \;\; \de = \frac{2}{3(1 + w_1)} \; .\label{a0 t} 
\er
Here $1/\tau = \ka \sqrt{3(1 + w_1) \, w_2 / 8}$, and $\tilde a_0$ is a normalization constant. At early times $a_0(t) \sim \tilde a_0 ( t / \tau )^{\de}$ the observed behaviour is typical  of  cosmologies  with  constant EoS: $p_0 / \vro_0 \approx w_1$. Later, for $t/\tau \gg 1$ the cosmological constant
\br
 \La_0 = \ka^2 w_2 / 8 (1 + w_1) \label{La0 e w2}
\er
dominates the EoS  and the universe enters a final, accelerated, asymptotically de Sitter phase, with an asymptotically constant energy density  $\vro_{\La_0} = w_2 / (1 + w_1)$. Then
\br
& a_0(t) \sim  \exp \left\{ 2 \sqrt{\frac{\La_0}{3}} \, (t - t_0)  \right\} \; . \label{La e tau}
\er
In particular, by choosing $w_1 = 0$, and thus $\de = 2/3$, we see that (\ref{EoS exemplo 2}) describes fairly well our observed Universe, neglecting inflation and the radiation dominated era: we begin at $t = 0$ with a dust-filled space-time, which ends at a de Sitter space with $\La_0$ presenting observed value \cite{Bousso_hepth_0702115}
\br
\La_0 \approx 3.14 \times 10^{-122} \times l_\Pl^{-2} \; ,
\er
 if we choose $w_2$ accordingly, using Eq.(\ref{La0 e w2}). 

\begin{figure}[htbp]
\begin{center}
\includegraphics[scale=0.8]{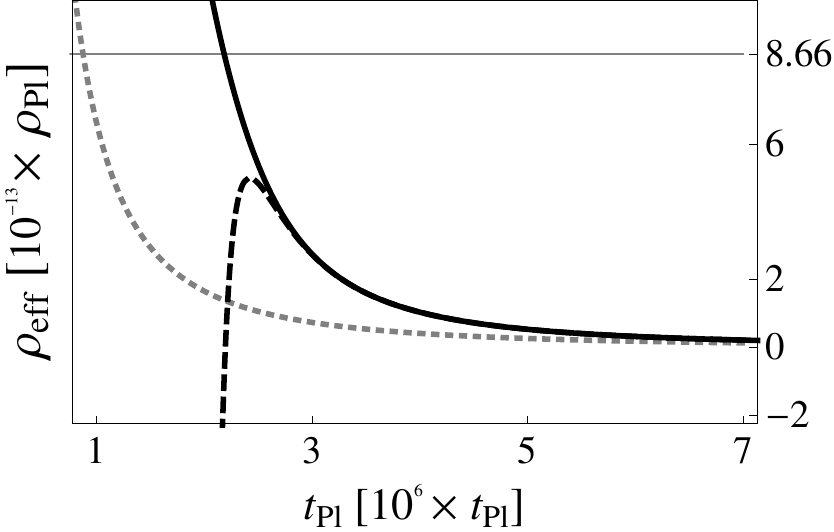} 
\caption{Time evolution of the energy density, for the same parameters of Fig.\ref{fig. EoS}. The black continuous curve gives the exact function $\vro(t)$, obtained from graphical inversion of $t(\vro)$ given by Eq.(\ref{t(H)}); the black dashed line gives the first order approximation; the dotted gray line depicts the EH density $\vro_0(t)$; the horizontal gray line marks the final energy density of the early acceleration period.}
\label{fig. vro de t}
\end{center}
\end{figure}

The first order corrections can be easily calculated from Eq.(\ref{before expansion}):
\br
&   H_{1}(t) =\frac{\tilde H_1}{ \sh^{2} \left( (t-t_0) / \tau \right) }\Big\{ \frac{t - t_0 }{4 \tau} - \frac{1}{8} \sh \left( \frac{2 (t - t_0)}{\tau} \right) -  \nonumber \\
& - \frac{1}{3} \coth \left(\frac{t - t_0}{ \tau} \right) \left[ \sh^{-2} \left( \frac{t-t_0}{\tau} \right) + \frac{5}{2} \right] \Big\}  ;	 \label{H1 t} \\
& A_1(t) = \frac{\tilde H_1\tau}{12}  \Big\{ \coth \left( \frac{t}{\tau} \right) \left[ 5 \, \coth\left( \frac{t}{\tau} \right) -\frac{3 t}{\tau} \right] +  \sh^{-4}\left( \frac{t }{\tau} \right) \Big\} \; , \label{A1 t}
\er
where $\tilde H_1 = - \ka^6 \, w_2^3 \, \tau L^4 / 72(1 + w_1)^2$. The constant $t_0 =  4 \pi \upsilon / 3(1+w_1) \m L^4$ assures that the singularity is placed in $t = 0$, and as $\m \to 0$ also $t_0 \to 0$.
From the scale factor and the Hubble function, one can determine all other relevant quantities, and in particular the effective energy density $\vro(t)$, plotted in Fig.\ref{fig. vro de t}.

As one approaches the initial singularity at $t = 0$, Eq.(\ref{t(H)}) shows that $H$ diverges. Eventually  we then have $H \gtrsim 1 / l_\Pl$ and the first order approximation is bound to fail -- indeed, $H_1(t)$ diverges more rapidly than $H_0(t)$ as $t \to 0$. This can be clearly seen in Fig.\ref{fig. vro de t}: the black continuous line shows the exact function $\vro(t)$, obtained from graphical inversion of $t(\vro)$ given by Eq.(\ref{t(H)}); the black dashed line shows the first order approximation $H_0(t) + \m H_1(t)$. It is clear that the first order approximation is only valid valid  for times greater than an instant $t_*$, when the curve has a maximum,  but for $t>t_*$   it is in good agreement with the exact solution. 
To smallest order in $|\m|$, we have
\br
t_* \approx \left( 2 (1 + w_1)^{3/8} + \pi \right) \frac{\sqrt2}{3 (1 + w_1)} \, L |\m|^{1/4} \; . \label{t* approx}
\er
Now, approximating to first order Eq.(\ref{cubic acc}) we find that the initial period ends when the effective energy density decreases to the value
\br
&\vro_{\rm{initial}} \approx \sqrt{\frac{1 + 3 w_1}{1 - w_1}} \,  l_\Pl^{-2} \;( L^4\, |\m|)^{-1/2} 	 ,			\label{vro acc init}
\er
which according to Eq.(\ref{t(H)}) happens at the instant
\br
t_\acc \approx  \frac{(\sqrt3 + \pi) \sqrt2}{3 (1 + w_1)} \; L \, |\m|^{1/4} \; .	\label{t acc approx}
\er
This shows that $t_*$ is slightly posterior to $t_\acc$, hence an approximation to only  first order is not sufficient to describe the initial acceleration period created by QTG.

As we have shown above, if $|\m| \gg |\m_\acc|$ we may have an eternally accelerated expansion. 
Such large values of $|\m|$, therefore, do not correspond reasonably to the observed universe. One might thus pose the question:  What  are \emph{the restrictions on the values of the gravitational couplings} $\la$ and $\mu$, which guarantee the physical consistency of the quasi-topological effects?
Regarding the initial acceleration period as an inflation era, we may assume that it would occur in the range of energies $\vro_\Pl \gtrsim \vro \gtrsim10^{-12} \times \vro_\Pl$ (see \cite{LindeInflationaryCosmology}), 
thus we must have $\m_\acc$ such that the root of (\ref{cubic acc}) lies within this bound.  The upper bound of this interval, viz. $\vro_\acc \sim 10^{-12} \times \vro_\Pl$, is indeed the case depicted in Figs.\ref{fig. EoS} and \ref{fig. vro de t}, what serves to demonstrate the validity of the first order approximation. 
Another phenomenological restriction  stems from the fact the the apparent horizon entropy should not be vanishing for too small values of $\vro$. 
In the GB case, there is no initial acceleration period, and this entropic restriction is in fact the only condition we have on  $\la$. It is quite evident from Eq.(\ref{vro GB}) that one may choose $\la$ to get $\vro_\GB$ as large as one needs -- e.g. for $\la = 3/4$ we have $\vro_\GB = \vro_\Pl$. In QTG instead, as it may be easily seen from Eq.(\ref{vro QTG}), for each  given  value of $\m$ determining the end of the acceleration period, we can choose $\la$ in order to place $\vro_\QTG$: \textit{(i)} before or \textit{(iii)} after $\vro_\ini$, or even to have \textit{(ii)} $\vro_\QTG = \vro_\ini$. This is exemplified in Fig.\ref{fig. EoS}, where the red lines represent the values of $\vro$ for which $s(\vro) = 0$. Note that in cases \textit{(ii)} and \textit{(iii)} the whole initial acceleration period is rendered ``unphysical'' on account of the negative horizon entropy there.


\section{On the late Universe QTG effects}

Although the more significant effects of QTG take place when the curvature, as well as the energy density, are big enough -- namely, at early times -- it turns out that the whole evolution of the universe is modified by the QTG terms. 
Clearly, at later times, as the curvature diminishes, the cubic and quadratic terms in (\ref{action non cf}) become more and more negligible and the first order approximation made in the last section is then justified. 

An example of such changes is given by the fact that the cosmological constant, which characterizes the geometry of the asymptotically $\dS_4$ spaces in the limit $t \to \infty$, is not equal to the bare one, $\La_0$, defined by  the matter Lagrangian. Indeed, in QTG one observes the \emph{effective} cosmological constant $\La_\eff$, related to the Hubble function $H(t)$. To first order in $\m$, one may determine $\La_\eff$  by  the approximation (\ref{H1 t}) for $H(t)$, or else by inverting directly the exact equation
\br
&\La_0 = \La_\eff \left( 1 - \m L^4 \La_\eff^2 /9 \right)  \; , 	\label{La eff e 0}
\er
obtained from Eq.(\ref{vro 0 eff}):  $\La_\eff \approx \La_0 + L^4 \m \, \La_0^3 / 9 \;$. 
This is in fact a general result, valid for every asymptotically de Sitter space, regardless of the particular  matter EoS leading to the final $\dS_4$ vacuum. Notice that for the very small value of the observed cosmological constant, both $\La_0$ and $\La_\eff$ are practically equal.

Another feature of asymptotically de Sitter space-times is the presence of a future event horizon.
Its comoving radius is given by the integral $r_f(t) = \int_t^\infty \dd t / a(t)$ and may be calculated to first order in $\m$  by  using the results (\ref{a0 t}) and (\ref{A1 t}). At zeroth order (i.e. in the EH case) this yields
\br
& r^{(0)}_f(t)   =   \frac{\tau}{\de \, \tilde a_0} \, \frac{1}{ \ch^\de(t / \tau)} \, \Hyp \left[ \frac{\de}{2} \, , \, \left(\frac{1 + \de}{2}\right) \, ; \, \frac{2 + \de}{2} \, ; \, \frac{1}{\ch^2(t/\tau)} \right]  , \label{rf EH}
\er
while the first order QTG correction is  given by:
\br
& r^{(1)}_f(t)  =  \frac{\tilde H_1 \tau^2}{12 \tilde a_0} \Bigg\{ 
\frac{5 / \de}{\ch^\de (t / \tau)} \, \Hyp \left[ \frac{\de}{2} , \frac{3 + \de}{2} ;  \frac{2+ \de}{2} ; \frac{1}{\ch^2(t/\tau)} \right] +   		\nonumber  \\
	& + \frac{1 / (4+ \de)}{[\ch(t / \tau)]^{(4 + \de)}} \, \Hyp \left[ \frac{4 + \de}{2} , \frac{5 + \de}{2} ;  \frac{6 + \de}{2} ; \frac{1}{\ch^2(t/\tau)} \right] -  \frac{3 t /(\de \tau)}{ \sh^\de(t  /  \tau)} - \nonumber  \\
	&   - \frac{3 / \de^2}{ \ch^\de(t / \tau)} \,  \Hyp \left[ \frac{\de}{2} , \frac{1 + \de}{2}  ;   \frac{2+ \de}{2}  ;  \frac{1}{\ch^2(t/\tau)} \right]   \Bigg\}   .	\label{r f correction QTG} 
\er
The function $r_f(t)$ describes the past light cone of the (infinite) future of the comoving observer at the origin. The tip of this cone is placed at $\{ r_f = 0 , \,  t = \infty \}$. 
Then  as $t \to \infty$  the \emph{physical radius}, $l_f (t)= a(t) \, r_f(t)$  becomes equal  to  the (constant) de Sitter radius $1 / H$.  The  first order approximation to $l_f(t)$ can be  easily  obtained with the  help  of Eq.(\ref{A1 t}): for example its value today, at $t = 13.8 \; {\rm{Gyrs}}$,  is  $l_f \approx \left( 4.90 \times 10^3  +  \m \times 9.10 \times 10^{-248} \right) {\rm{Mpc}} \; $.

Alternatively, the comoving radius $r_p(t) = \int^t_{\tilde t} \dd t / a(t)$ describes the future light cone for an observer at the origin, starting from its tip at $\{ r_p = 0, \, t = \tilde t \}$. In a singular universe, if this tip is placed at the beginning of time, $\tilde t  = 0$, then $r_p(t)$  is  the particle horizon.  In practice, the choice of $\tilde t$ determines only a constant, for we may write
 \br
\hspace{-0.7 cm} r_p(t) = - \int_t^\infty \dd t / a(t) + \int_{\tilde t}^\infty \dd t / a = - r_f (t) + r_f (\tilde t) \; . \label{r p e f}
 \er
Thus for example, in the EH case with $\tilde t  = 0$, Eq.(\ref{rf EH}) gives
\br
\hspace{-1 cm} & r^{(0)}_p(t)   = -  \frac{\tau / \de \tilde a_0}{ \ch^{\de}(t / \tau)} \, \Hyp \left[ \frac{\de}{2} ,  \frac{1 + \de}{2}  ;  \frac{2 + \de}{2}  ; \frac{1}{\ch^2(t/\tau)} \right]  + \frac{\tau}{2 \tilde a_0} \Bet \left(  \frac{\de}{2} , \frac{1 - \de}{2} \right) ,    \label{rp 0}
\er
and in general Eqs.(\ref{rf EH}) and (\ref{r f correction QTG}) determine also the first order correction for $r_p(t)$, for a given $\tilde t$.

 Recall that  the first order approximation for the scale factor is only valid for $t \gtrsim t_*$, with $t_*$ denoting the instant where the approximation fails, Eq.(\ref{t* approx}). Therefore in QTG we have the ``technical'' impossibility of placing the tip of the light cone on the initial singularity -- the earlier we may place it is at $\tilde t = t_*$. 
Due to the fact that $t_*$ (at our first  approximation) is  slightly posterior to $t_\acc$, we are  then technically prevented from describing the increase of the particle horizon during this early ``inflationary'' period. Consequently, we cannot determine its number of $e$-foldings, and whether it solves the usual problems of non-inflationary cosmology -- such as, for example, the horizon problem. 

The knowledge of the above  expressions for the radii $r_p(t)$ and  $r_f(t)$ provides an analytic description of the \emph{causal diamond} \cite{Gibbons1, BoussoNbound} and its comoving volume $V_{\cd}(t)$, in the QTG cosmological model under investigation. The former is defined as the intersection of the causal past of the ``point'' $\{ r_f = 0 , \,  t = \infty \}$ and the causal future of  $\{ r_p = 0, \, t = \tilde t \}$. At each instant $t$ the volume is given by $V_\cd=\frac{4}{3} \pi r^3(t)$ with $r=r_p$ on the upper and $r=r_f$ on the lower half of it. It is  easily  seen  that $V_\cd$ has a single maximum at $t=t_\edge$, when $r_p (t_\edge)=r_f (t_\edge)$ \cite{BoussoNbound}. With the aid of Eq.(\ref{r p e f}), the instant $t_\edge$ is determined from the equation $r_f(t_\edge) = \frac{1}{2} r_f(\tilde t)$, which is  exact, i.e. independent from the first order approximation. If $r_f(t)$ is a continuous function, then we conclude that a first order correction to $\tilde t$ must imply a first order change in $t_\edge$. Therefore for small $t_* = \tilde t$ in QTG, the edge of the causal diamond occurs at an instant differing  not more than at first order from the corresponding value in EH.

 In EH gravity with linear EoS (\ref{EoS exemplo 2}) and $w_1 = 0$, describing  an early universe dominated by dust\footnote{In which not only the initial inflationary period, but also the radiation-dominated era of the concordance model are absent.}, Bousso et al. \cite{Bousso_hepth_0702115} have calculated $V_\cd (t)$ and used it to predict the order of magnitude of the observed cosmological constant. Their analysis, based on the rather universal (phenomenological) evaluation of the entropy production rate in our universe, demonstrates that only when $\La_0 \sim 10^{-122} \times l_\Pl^{-2}$ \emph{the causal entropic principle (CEP)}, requiring maximal entropic production within the corresponding causal diamond volume, is fulfilled. According to their arguments the main production of bulk entropy occurs during the matter-dominated era. Therefore one can perform a similar analysis for the QTG extension of this linear EoS model, by imposing the  $CEP$ conditions for the causal diamond whose inferior tip is placed at $\tilde t = t_*$, thus respecting the restrictions of the considered first approximation. Surely, even if the tips of the causal diamonds in QTG and in EH gravity were placed at the same point -- say, by replacing $t = 0$ with $t = t_*$ in the EH case as well --, still they would not be identical due to the changes in the dynamics given in (\ref{r f correction QTG}). However, as we saw above, the  first order QTG corrections we are considering do not change significantly the comoving volume of the causal diamond, and in fact they lead to the same prediction for the  magnitude of the effective cosmological constant  $\La_\eff \approx \La_0 + L^4 \m \, \La_0^3 / 9 \;\sim 10^{-122} \times l_\Pl^{-2}$.

Throughout this discussion, we have not taken into account the entropic  bounds  derived in Sect.\ref{sect. horizon entropy}. Regardless of our technical restrictions for placing the inferior tip of the causal diamond, in the case of  $\m < 0$,  we cannot place it before the  instant  $t_\QTG$ when the apparent horizon's entropy vanishes. The same is true in the GB case, for $\la > 0$. We are however assuming that $t_\QTG$ is very small, in particular that $t_\QTG \lesssim t_*$. This can always be set for an appropriate value of $\la$. Either way, the entropic restrictions \emph{do not allow} the inferior tip of the causal diamond to be placed on the singularity.

Let us list in conclusion a few open problems concerning the considered (linear EoS) cosmological model of the cubic Quasi-Topological gravity: (a)  The stability conditions for the asymptotic $\dS_4$ cosmological QTG solutions, which requires the calculation of the spectrum of the corresponding linear fluctuations at least in the probe approximation; (b) The analysis of the properties of QTG models with more realistic matter content by considering EoS or equivalently (string inspired) matter superpotentials \cite{KKLT}  giving rise to early-time  inflation with  a \emph{desired slow-roll behaviour}, which is under investigation \cite{LoveCosmo_tobepublished}. It is worthwhile to also mention that the methods and some of the results of the present paper seem to have a straightforward ``cosmological'' application to the recently constructed ``higher curvature'' quartic QTG extension \cite{mann} of the EH gravity, as well as to the case of  \emph{spatially curved FRW}  solutions of the considered  four dimensional,  cubic QTG model.

\textit{Acknowledgments}. We are grateful to C.P. Constantinidis for his collaboration in the initial stage of this work and for the discussions.






\bibliographystyle{model3-num-names}
\bibliography{<your-bib-database>}



\end{document}